# Implementation of password manager with SRAM-based Physical Unclonable Function

Mohammad Mohammadinodoushan

Northern Arizona University, mm3845@nau.edu

## ABSTRACT

Hacking password databases is one of the most frequently reported cyber-attacks. Current password management systems are based on known and public algorithms. Also, many studies have shown that users select weak passwords. Thus, with the emergence of new powerful computing devices, the passwords based on known algorithms can be disclosed. Using physical unclonable functions (PUFs) for increasing the security level of password management systems is a quite recent method that is proposed to solve this problem. In this method, Addressable PUF Generator (APG) is added to the conventional password management systems. This report is aimed at implementing the password generation scheme using SRAM-based PUF. The bit error is indeed the main issue with using PUFs is addresses in this paper. To solve this issue, Ternary Addresseble PUF Generator is used.

# 1 Introduction

Traditional password (PW) management systems have used the mapping of passwords that were based on the known and public algorithms like hashing and salting [1]. However, research shows that most of the users choose common and weak passwords [2-5]. Thus, with the emergence of new powerful GPUs and computing devices, the PWs can be disclosed by brute force attacks. Another method that has been used in the literature is encrypting a PW with a key saving them in a database. However, the problem with this method is that when the key is disclosed, the PWs are also disclosed. Several systems and methods have been proposed in the literature on using physical unclonable functions (PUFs) for password generator systems to improve the level of security. In this method, Addressable PUF Generator (APG) works as a hardware protection layer. This hardware layer can be added to the software layer that can is commonly used in the PW generator systems. Using APG leads to both cost reduction and security improvement. The organization of the remainder of this document is: Section 2 contains background information about Ternary SRAM PUF, and hash-based PW management/generation schemes. Section 3 includes detailed information about the system design of including the protocol. Part 4 will focus on implementation and the results.

# 2 Background

## 2.1 PUF Designs

Many forms of PUFs have been designed in the past, which are reviewed . One type of PUFs is memory-based PUF like SRAM [6, 7] or MRAM PUF [8]. PUFs have been used in many cryptographic applications such as Password Management Systems [9, 10], key exchange [11], Keyless encryption [12], and Power Storage [13-17]. There is an increasing interest in memory-based PUFs since it is possible to make them from memories already existing in many systems. The other advantage of this kind of PUFs over conventional PUFs, is their size and speed, while they need less power comparing to traditional PUFs. Besides, the size of the required PUF is insignificant comparing to the size of the whole memory. Thus, the location of the used PUF can be another secret information, which increases security, when a hacker gets access to the memory. Using memory cells for designing PUFs in security protocol has been used a lot in the previous bodies of research.

### 2.1.1 SRAM-PUF

Static random-access memory (SRAM) PUFs were discovered, based on the original SRAMs independently and concurrently, by Holcomb [18]et al. [18, 19] and Guajardo et al. [20]. As shown in **Fig. 1**, a typical SRAM cell contains two cross-coupled inverters at its core with two stable states. The random physical mismatch between two is controlled by the power-up behavior, which itself is determined by the manufacturing variability. During power-up, some cells prefer storing 0, and some prefer storing 1. Responses of a PUF are the same in different cycles for these kinds of cells [21, 22].

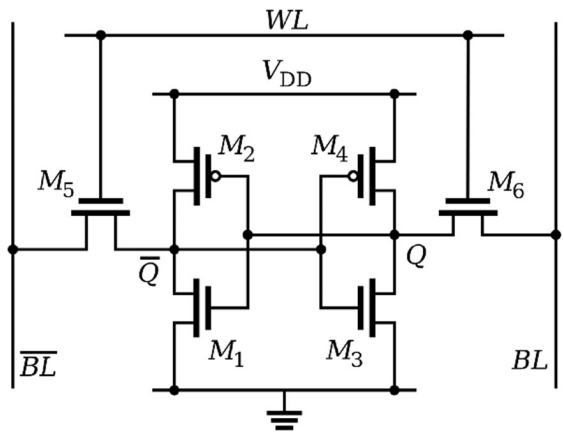

**Fig. 1 SRAM Cell, Source: Adapted from [18-23]**

## 2.2 Addressable PUF Generator (APG)

The structure of Addressable PUF Generator (APG) [24, 25] is shown in **Fig. 2**. In this architecture, the controller sends an address XY to the memory array to read cells from which challenges and responses are generated. As described in [24, 25], APG can be used to create temporary passwords and authenticate users.

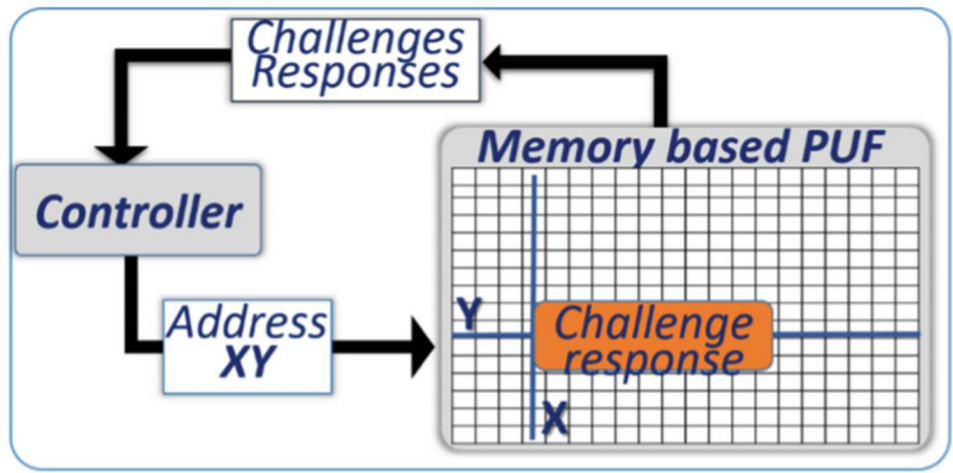

Fig. 2 Block Diagram of the APG, Source [26]

## 2.3 Ternary PUFs

The idea of ternary PUF is to select only the stable cells to generate the PUF output and ignore the other cells. In this section, the method of identifying fuzzy cells in both ternary SRAM and MRAM PUFs are discussed.

### 2.3.1 Ternary SRAM PUF

As mentioned before, the method to generate PUFs from SRAM arrays is to subject the device to power-off power-on cycles. A significant proportion of the cells always is read as either "0" or a "1". However, a few numbers of SRAM cells have a weak preference or no particular preference at all [26]. Thus, the responses of the PUF is not stable for these cells, which are known as fuzzy cells. The number of fuzzy bits divided by the total number of bits in the pattern is defined as the SRAM PUF noise [27]. An intuitive approach to solve the issue with fuzzy cells is to read the PUF cells repeatedly and chose only those cells that always provide the same output. In the characterizing (enrollment) phase, the SRAM-PUF cells are read hundreds of times by repeating the power-off power-on cycles. In each cycle of reading, the unstable cells are specified that have different responses against their previous response and denoted by an "$X$" mark. In this way, it is possible to recognize the cells that can generate stable "0" and "1" and remove those with the "$X$". The higher number of reads is accompanied by creating a more stable challenge. From now on, only stable PUF cells generate a stable response [10].

# 3 Protocol

This report focuses mainly on explaining the implantation of the PW generator with APG shown in **Fig. 3**.

The architecture input is user ID (ID hereafter) and Password (PW hereafter) of the user. The main output of APG that is shown in **Fig. 3** is the PUF response [10]. The response of PUF is generated through the following steps:

Step 1- ID and PW are the input of Addressable PUF Generator. Message Digest (MD) of PW is calculated in microcontroller (MCU) with software SHA2-256.

Step 2- The MD of the PW is the input of the "Expander block" in **Fig. 3**. The details about the Expander block are shown in **Fig. 4**. The output of the hash (or the input of Expander block) is 32 bytes, 256 bits. Since the size of the used SRAM PUF is 8 kilobytes, 16 bits are needed for addressing a specific SRAM cell. Also, the input of the Expander block, which is 256 bits, can only point to 16 independent addresses. Therefore, the extracted challenge would be 16 bit, which ends with low entropy. Thus, we cannot cover enough number of users. Expander block is designed for increasing the entropy. Increasing the entropy is done by generating longer MD from the original MD. As can be seen in **Fig. 4**, the left two bytes of the MD are rotated eight times. Then, all the results of shifting MD will be the inputs of SHA2-256, as shown in **Fig. 4**. The longer message digest is 256 bytes, which are the concatenation of 8 message digests. These 256 are enough to generate a 128-bit challenge/response stream and provide high entropy.

Step 3- The outputs of "Expander block" are 128 addressees of SRAM cells. These addresses include the address of fuzzy cells

Step 4- The outputs of "Expander block" is revised in the Masking block (Masking fuzzy cells). In this block, the addresses of the fuzzy cells are ignored. The next non-fuzzy cell after each fuzzy cell is considered for extracting the challenge or response of PUF. The output of the masking block is 128 addresses.

Step 5- Each bit of PUF response is obtained in 128 addresses generated in Step 4. Therefore, the response of PUF is 128 bit, as shown in **Fig. 3**.

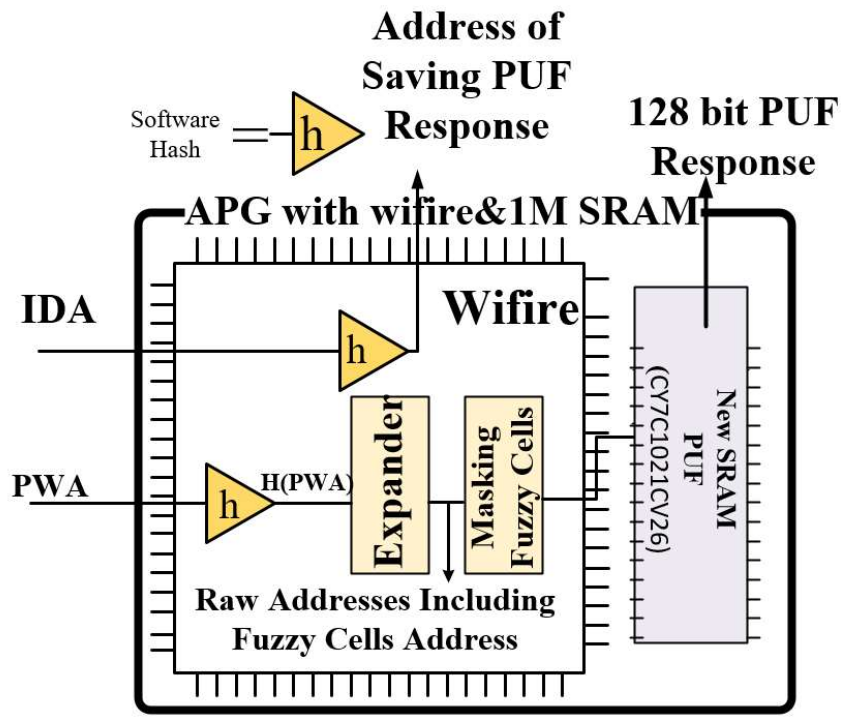

**Fig. 3 The implemented architecture for APG with wifire and 1Mbit Cypress SRAM**

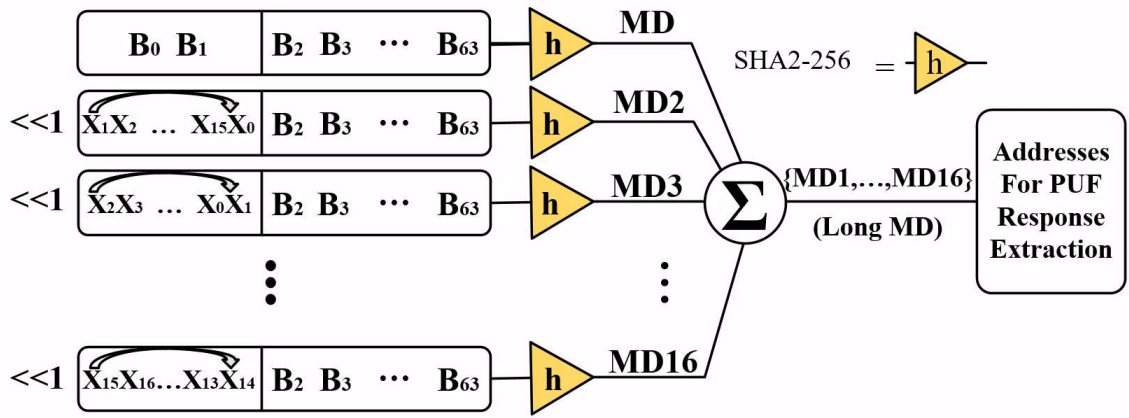

Fig. 4　Details of "Expander block" shown in Fig. 3 [10]

# 4 Implementation and results

The APG is implemented using WiFire development boards from Digilent, powered by Microchip and 1-MBIT CYPRESS SRAM[28], as shown in **Fig. 5**. The SRAM part number is CY7C1021CV26, which is a 1Mbit SRAM arranged in 64 K words every 16 bits. A similar protocol [10] for SAMV71 MCU with Cortex M7 core and 32 Kbyte SRAM is revised to make it compatible with wifire board. The coding environment for coding is Arduino IDE due to its simplicity. The functions used in the code and the result of each function are explained briefly in the next coming sections.

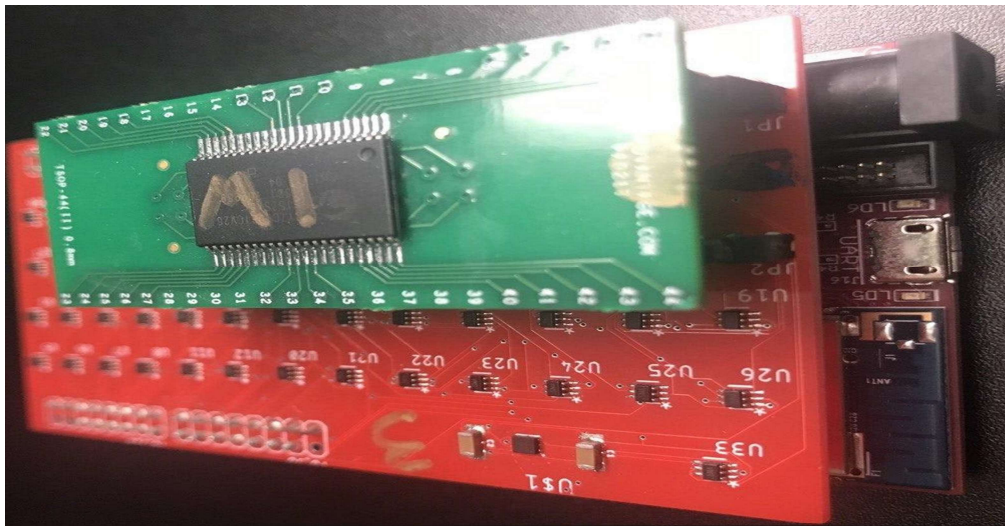

Fig. 5 APG device developed by NAU cyber security Lab

## 4.1 Receiving data

At the beginning of the program, the ID and password (PW) is sent to microcontroller (MCU) which is WiFire board. MCU captures and echo every single received byte and then the ID and PW in the HEX format to the terminal. The codes in **Fig. 6** shows how the data is received, echo back and printed in hex format for ID string:

```
  Serial.print("\n\r\n\r Enter your ID\n\r");
  recv_char = 0;
  i_counter = 0;
  while ((recv_char != '\n') && (recv_char != '\r') && (i_counter < MSG_SIZE))
  {
   if(Serial.available())
    {
     recv_char = Serial.read();
     Serial.write(recv_char);        // echo to terminal
     id_buff[i_counter++] = recv_char; // queue character into buffer
    }
  }
  id_buff[i_counter-1] = '\0';        // the last character was enter, then replace it with NULL
  id_size = i_counter-1;
```

Fig. 6 Code used for receiving data of the user

The result of this step is shown in **Fig. 7**

```
 Enter your ID
PasswordManagementWithWifire

 Enter your Password
1-MBIT SRAM

Entered ID:
0X50 61 73 73 77 6F 72 64 4D 61 6E 61 67 65 6D 65 6E 74 57 69 74 68 57 69 66 69 72 65
Entered Password:
0X31 2D 4D 42 49 54 20 53 52 41 4D
```

**Fig. 7 Results of receiving and printing the ID and PW**

## 4.2 Processing data and making the address

The results of steps that occur in the "Expander block" are shown in **Fig. 8** and **Fig. 9**. These results are obtained after entering "PasswordManagementWithWifire" and "1-Mbit SRAM" as the "Username" and "Password". As previously mentioned, the output of the hash digest (or the input of the Expander block in **Fig. 3**) is 32 bytes. As can be seen in **Fig. 4**, the two most significant bytes of the MD is rotated 8 times. Then, all the results of shifting MD will be the inputs of SHA2-256, as shown in **Fig. 4**. The results of these eight messages digest are shown in **Fig. 9**.

```
Results of Shifting Message Digest:
16 A8 DE D1 42 09 57 EB 04 2B C4 0F 08 DB 2B 2C 14 52 BB 4D E6 E0 63 B5 ED 43 9C 26 7C 15 78 04
2D 50 DE D1 42 09 57 EB 04 2B C4 0F 08 DB 2B 2C 14 52 BB 4D E6 E0 63 B5 ED 43 9C 26 7C 15 78 04
5A A0 DE D1 42 09 57 EB 04 2B C4 0F 08 DB 2B 2C 14 52 BB 4D E6 E0 63 B5 ED 43 9C 26 7C 15 78 04
B5 40 DE D1 42 09 57 EB 04 2B C4 0F 08 DB 2B 2C 14 52 BB 4D E6 E0 63 B5 ED 43 9C 26 7C 15 78 04
6A 81 DE D1 42 09 57 EB 04 2B C4 0F 08 DB 2B 2C 14 52 BB 4D E6 E0 63 B5 ED 43 9C 26 7C 15 78 04
D5 02 DE D1 42 09 57 EB 04 2B C4 0F 08 DB 2B 2C 14 52 BB 4D E6 E0 63 B5 ED 43 9C 26 7C 15 78 04
AA 05 DE D1 42 09 57 EB 04 2B C4 0F 08 DB 2B 2C 14 52 BB 4D E6 E0 63 B5 ED 43 9C 26 7C 15 78 04
54 0B DE D1 42 09 57 EB 04 2B C4 0F 08 DB 2B 2C 14 52 BB 4D E6 E0 63 B5 ED 43 9C 26 7C 15 78 04
```

**Fig. 8 Rotating the first two bytes of the password MD**

```
8 MD results:
MD1: 16 A8 DE D1 42 09 57 EB 04 2B C4 0F 08 DB 2B 2C 14 52 BB 4D E6 E0 63 B5 ED 43 9C 26 7C 15 78 04
MD2: 0C 16 CB EB 67 25 7E D2 73 0D 77 88 23 76 B2 F2 BF 8A 1C 8A 47 06 60 6E BA 02 F3 86 80 AB EC 53
MD3: 31 4F 1C 02 14 7F 0F EC EB 79 D9 FE F4 AC 64 5A 25 08 7A 56 86 DE 5A C5 97 2A 4D 44 19 83 D3 B7
MD4: 1C 61 95 65 06 47 EB 2B 6E 30 AE 4F 95 D5 C7 41 90 89 C0 14 EE D3 F2 7E D9 37 BA 48 C8 C8 2E 3F
MD5: D3 DD 4E 97 0F A6 FA C2 9E 7A F5 21 01 99 43 46 64 51 73 B1 37 1C 8D 62 E3 F3 A5 E0 E8 43 E3 A6
MD6: DC F1 43 A4 B2 8E 01 6B 63 22 5A 54 D7 B5 CC 5F 26 3A AE 4C 4E 7F FA 1A E1 7C 49 9D BF E4 B3 4E
MD7: E3 8F BC EB 70 4F 6D 8A 16 CD CB 70 1B E3 31 AF E3 BF 59 79 33 17 94 B3 61 D8 B7 AF 92 34 C9 87
MD8: 05 A7 10 72 12 E9 D4 10 58 D2 5B 62 76 76 A8 69 53 4B 97 85 3B 47 86 E2 2D 6E CA 8D 26 97 2C CE
```

**Fig. 9 Message digests from hashing of shifted results**

Finally, 8 MD (each MD is 32 Bytes) results that are shown in **Fig. 9** will be concatenated to generate long MD in Expander block. Therefore, the length of Long MD is 8*32 = 256 Bytes, which is used to generated "128 Addresses for extracting PUF Response". Since just 8 kilobytes of SRAM is used as PUF, we need 16

(2 bytes) bits for addressing a specific cell. Thus, we will have 256/2=128 addresses, which are shown in **Fig. 10**.

```
128 Addresses for extracting PUF Response:
01CF 66BD 987D 5DE4 3108 BC60 FDE9 4C68 16EC 137B CF4B EFFE FA8F EF98 495F 2DBE
DE3A 3929 C17A 835C A862 B6CF 406F 782F 4111 E94A 187F 332D 6B1F A7D7 3E44 0A1E
7583 B383 9642 EE24 67B9 4095 8FDB 55D8 4441 BB31 18E6 1C4C 25EB F2E4 A1D8 1079
406D 0D64 E4D6 0274 ECBE 6589 567E F2C2 CEC8 2812 027E 92DC E309 E7C9 0DE7 F56A
357C A0ED 861F D702 B059 B0FC 7C3F 233D E7AD CBE0 02E4 EBD6 EFC1 04DA F24B 86ED
C82F F9B3 2BC2 832A 833F 5D22 C026 68D8 9A79 6FA1 1F2E 6376 9E5D 4C0A 0C1F 6F14
AF60 2A55 ECB1 A85E 9389 2749 367D B9A1 111A 3E16 DAC1 993E 8424 2016 846E AABD
7D25 E72B 560A 0D92 F049 8DE1 AD7B 93B3 2C40 44CB D8E8 6DFA 33AE 61E5 3ACC 517E
```

Fig. 10 Generating "128 Addresses for extracting PUF Response" after Masking the fuzzy cells

Each bit of the challenge/Response is extracted in 128 addresses as shown in Fig. eee. The extracted SRAM-PUF response is shown in **Fig. 11**.

```
128bit PUF Response:
00000101010010011011101011001000101001101010000100000111001110100010000011101101
1101100010111010010011000001101001001100010000110
```

Fig. 11 128 bit PUF response extracted in 128 addresses shown in Fig. 10

**Acknowledgments**

The author is thanking the contribution of students at the cybersecurity lab at Northern Arizona University, in particular Ian Burke and Christopher Philabaum